\documentclass[iop]{emulateapj}
\usepackage{natbib,graphicx}
\usepackage{tabular}
\usepackage{color}
\usepackage[dvipsnames]{xcolor}

\begin{document}

\submitted{Submitted to the Astrophysical Journal}
\journalinfo{Submitted to the Astrophysical Journal}
\title{A Radio Emission Analysis of Classical Nova V351~Pup (1991)}

\author{Carolyn Wendeln$^{1,2,3}$, Laura Chomiuk$^{2}$,Thomas Finzell$^{2}$, Justin D. Linford$^{2}$, Jay Strader$^{2}$}
\altaffiltext{1}{Department of Climate and Space Sciences and Engineering, University of Michigan, Ann Arbor, MI 48109, USA}
\altaffiltext{2}{Department of Physics and Astronomy, Michigan State University, East Lansing, MI 48824, USA}
\altaffiltext{3}{Lyman Briggs College, Michigan State University, East Lansing, MI 48824, USA}

\begin{abstract}

Previously, Nova Puppis 1991 (V351 Pup) was measured to host one of the most massive ejections claimed in the literature. Multi-frequency radio detections from one epoch were published for this nova in the 1990's; and yet, the remaining data collected by the Karl G. Jansky Very Large Array (VLA) have remained unpublished. In this paper, we analyze the remaining unpublished data sets for V351 Pup at frequencies of 4.9, 8.4, 14.9, and 22.5 GHz.
We fit the resulting light curve to a model of expanding thermal ejecta, under the assumption that the radio emission is dominated by free-free radiation and accounting for high levels of clumping in the ejecta.
Images of V351 Pup in both the radio (from the VLA) and H$\alpha$+[\ion{N}{2}] (from \emph{Hubble Space Telescope}) exhibit no aspherical structure, strengthening our assumption of spherical symmetry. From expansion parallax methods, we estimate the distance to V351~Pup to be $5.0 \pm 1.5$ kpc.
Our light-curve fit yields a value of $log_{10}(M_{ej})={-5.2} \pm {0.7}$ M$_{\odot}$ for the ejecta mass, implying that V351 Pup is on the low end of expectations for ejecta mass from classical novae. A comparison between our derived ejecta mass and theoretical models gives evidence for a very massive (1.25 M$_{\odot}$) white dwarf, which is consistent with spectroscopic evidence for an oxygen-neon white dwarf.
\end{abstract}


\section{Introduction}\label{intro}

A classical nova outburst is a thermonuclear runaway that occurs on the surface of a white dwarf (WD) in a close binary. The Roche-lobe-filling secondary transfers hydrogen-rich gas to its companion WD via an accretion disk \citep{1976MNRAS.176...53G}.
The accreted material gradually forms layers of fuel as it falls onto the surface of the WD. The resulting layers are compressed and heated by the strong surface gravity of the WD \citep{Starrfield}. The more massive the WD, the smaller its radius and surface area, and less accreted mass is needed to reach to the pressure and density required for a TNR \citep{2005ApJ...623..398Y}. The thermonuclear runaway begins when the bottom of the accreted layers reaches $\sim$$10$ million K, which corresponds to the critical pressure and temperature sufficient for CNO cycle burning of hydrogen \citep{1986ApJ...310..222P}.

The thermonuclear flash powers an optically thick wind that ejects nearly all of the accreted material \citep{2000NewAR..44...81S}. The eruption ejects matter on the order of $ 10^{-7}$ to $10^{-3}$ M$_{\odot}$, at velocities ranging from hundreds to thousands of km s$^{-1}$ \citep{1998PASP..110....3G,2005ApJ...623..398Y}. Unlike a Type Ia supernova, the WD is not destroyed during the outburst and can continue to accrete material from its companion after eruption. As long as accretion continues, all novae are expected to recur on timescales from $\sim$$10^{8}$ years to less than a year \citep[e.g.][]{2005ApJ...623..398Y}.

Properties of the binary star system, such as the WD mass and accretion rate from the companion star, dictate the parameters of the nova explosion. Such parameters include the recurrence time, ejecta mass, and explosion energetics \citep[e.g.][]{2004RMxAC..20..192T, 2005ApJ...623..398Y}. Theory does a fair job at predicting these simple relationships between the WD properties and explosion characteristics \citep{2005ApJ...623..398Y,2013ApJ...777..136W}; however, these predicted relationships have been hard to test observationally. There is an order of magnitude discrepancy between observational and theoretical predictions in ejecta masses \citep{1998MNRAS.296..502S, 2000NewAR..44...81S, 2012BASI...40..293R}.

Nova Puppis 1991 (V351 Pup) was discovered in outburst after optical maximum on 1991 December 27 \citep{1992IAUC.5422....1C}. While \citet{1996MNRAS.279..280S} suggested its companion might be a red giant, the 2.8 hr orbital period is more consistent with a main-sequence donor \citep{2001MNRAS.328..159W}. It was a relatively fast nova with $t_{2}$ (the time for the optical light curve to decline by two magnitudes) estimated to be between 10 and 16 days \citep{1995ASSL..205..303P,1996MNRAS.279..280S,2000AJ....120.2007D}. Abundance estimates from UV/optical spectroscopy imply that the primary was an ONe type WD \citep{1996MNRAS.279..280S}. Published estimates for the distance to the nova range from 2.7--4.7 kpc \citep{1996ASPC...93..174H,1996ApJ...466..410O,1996MNRAS.279..280S,2000AJ....120.2007D}. Measurements for the maximum velocity of the expanding shell range from 2500 to 3200 km s$^{-1}$ \citep{1992IAUC.5427....1D,1992IAUC.5428....2S, 1994ApJS...90..297W}

V351 Pup was previously measured by R.\ Hjellming to host one of the most massive ejections claimed in the literature \citep{1996ASPC...93..174H, 1998PASP..110....3G,Seaquist_Bode08}.
The ejecta mass of V351 Pup was listed to be $1.01 \times 10^{-3}$ M$_{\odot}$ from radio frequency observations; however, the light curve and details of the ejecta mass derivation were not presented, so it is difficult to assess them in comparison with our results below \citep{1996ASPC...93..174H}. In contrast, \citet{1996MNRAS.279..280S} reported a shell mass of $2.0 \times 10^{-7}$ M$_{\odot}$ based on UV/optical spectroscopy. A large fraction of this discrepancy is due to Hjellming's assumption of no clumping in the nova ejecta, corresponding to a filling factor of unity, while Saizar et al. (1996) found a volume filling factor of $1.5 \times 10^{-5}$.

Sixteen months after V351 Pup's optical maximum, hard X-rays were detected by \emph{ROSAT} without the accompaniment of a soft X-ray component \citep{1996ApJ...466..410O}. These hard X-rays in novae are usually thought to come from shocks, though, in the case of V351 Pup, magnetically controlled accretion was proposed as another possible source \citep{2001ApJ...551.1024M}. Soft thermal X-rays are associated with nuclear burning on the WD \citep{2005AA...439.1061S, 2011ApJS..197...31S}.
It has been speculated that the presence of hard X-rays from V351 Pup could have been caused by either shocks in the ejected shell or by magnetically controlled accretion onto the WD \citep{1996ApJ...466..410O}.
\citet{2001MNRAS.328..159W} noted that their quiescent optical light curve for V351 Pup has a strong resemblance to the intermediate polar V1500 Cyg, which hosted a nova explosion in 1975. The assertion of a magnetic nova was fairly speculative and has yet to be confirmed.

In this paper, we present the radio observations of the 1991 outburst of V351~Pup, including a multi-frequency light curve and spatially resolved imaging (complemented by \emph{Hubble Space Telescope (HST)} imaging of the nova ejecta). In \S 2, we discuss the standard model for radio emission from novae, and we present the radio observations of V351~Pup in \S 3. We present the radio and optical imaging of the nova shell in \S 4, and use these images to derive a distance, via expansion parallax, to V351~Pup. We fit the radio light curve in \S 5, and discuss implications in \S 6, to conclude in \S 7.

\section{Radio Emission and Model Fitting}

\subsection{\textit{Radio Emission}}

Observations of novae at radio frequencies give insight into the total ejected mass, density profile, and the kinetic energy of a nova eruption. Radio frequencies primarily trace the warm ionized gas in nova ejecta. In most cases, the radio emission from novae is dominated by thermal free-free emission \citep{Seaquist_Bode08}. Radio observations have proven to be beneficial to our understanding of novae because the external and/or internal extinction by dust is negligible.
The characteristic time-scale of radio light curves for nova eruptions is several years after optical maximum.
The emitting gas remains optically thick at radio wavelengths for a prolonged time, causing the maximum radio flux density to occur much later in time than optical maximum \citep{Seaquist_Bode08}.

At the beginning of the eruption, the shell of ejected matter is optically thick and opaque at all radio frequencies \citep{Seaquist_Bode08}. The
emitting surface area expands with the ejecta, resulting in an increase in the radio flux. The flux density at this early time rise depends on the distance to the nova, the electron temperature, and the expansion velocity of the radio photosphere \citep{1977ApJ...217..781S,1979AJ.....84.1619H,1980AJ.....85..283S}. The photoionization heating of the expanding ejecta by the hot WD maintains the observed, nearly constant, warm ($\sim 10^4$ K) plasma during this phase \citep{2015ApJ...803...76C}. During this rising phase of the radio light curve, the radio spectral index $\alpha$ (defined as $S_{\nu} \propto \nu^{\alpha}$), approaches $\alpha=2$ as expected
for optically thick thermal emission \citep{Seaquist_Bode08}.

As time passes, the expanding shell becomes less dense, less optically thick, and the radio photosphere lags behind the expansion of the outer shell boundary \citep{Seaquist_Bode08}. The shell becomes optically thin at higher frequencies first, with lower frequencies becoming optically thin later in time \citep{1977ApJ...217..781S,1979AJ.....84.1619H,1980AJ.....85..283S}. When the ejected mass is optically thin at all frequencies, the spectral index becomes nearly flat, reaching $\alpha$=$-0.1$ \citep{Seaquist_Bode08}.

Modeling of the thermal bremsstrahlung emission provides us with the total ejected mass, density profile, and the kinetic energy, once the ejection speeds of material and distance to the nova are known \citep{Seaquist_Bode08}. Furthermore, this form of modeling has a weak dependence on the temperature of the ejecta. Such modeling of thermal emission from an expanding shell has been used to explain radio light curves from $\sim$10 novae in the past \citep{1977ApJ...217..781S,1979AJ.....84.1619H,1980AJ.....85..283S,1983MNRAS.202.1149K, 2012BASI...40..293R}.
Radio light-curve model fitting is derived assuming a finite shell undergoing homologous expansion (Hubble flow), a wind with constant velocity and a variable mass-loss rate (variable wind), or a combination of properties between the two models known as the ``unified model" \citep{1996ASPC...93..174H, Seaquist_Bode08}. The predicted light curves from these models assume a similar density profile ($\rho \propto r^{-2}$ or $\propto r^{-3}$), spherical isothermal ejecta, and unity filling factor. The differences in the models lie in their treatment of the inner boundary of the ejecta, which determines how the radio light curve becomes optically thin. More recent work has tweaked the geometry of this model, with the addition of either bipolar outflows \citep{Ribeiro} or clumping \citep{2014ApJ...785...78N}.

\subsection{\textit{Hubble Flow Model for Mass Ejection}}

In this paper, we fit the radio light curve of V351 Pup to the Hubble flow model because the Hubble flow produces results similar to the variable wind model but with fewer free parameters. The Hubble flow model describes a homologously expanding ionized shell, where the velocity gradient increases linearly with radius ($v \propto R$; \citealt{1979AJ.....84.1619H, 1996ASPC...93..174H}). The radio luminosity at early times, when the shell is optically thick, is proportional to $R_{\rm out}^{2} \times T_e$, where $R_{\rm out}$ is the expanding outer radius of the nova ejecta and the electron temperature, $T_e$, is held constant as a function of time and radius \citep{Seaquist_Bode08}.

The Hubble flow model assumes that the finite shell is ejected at time $t_0$, with inner and outer radii obeying the following equations:
\begin{equation}
\ R_{\rm in} = v_{\rm in}(t-t_{0})+R_{{\rm in},0}\
\end{equation}
\begin{equation}
\ R_{\rm out} = v_{\rm out}(t-t_{0})+R_{{\rm out},0}\
\end{equation}
where $R_{\rm in}$ is the innermost radius at an arbitrary time $t$, $R_{\rm out}$ is the outermost radius at $t$, $R_{{\rm in},0}$ and $R_{{\rm out},0}$ are the initial innermost and outermost radii, respectively, at $t_0$, $v_{\rm in}$ is the minimum velocity of ejected material, and $v_{\rm out}$ is the maximum velocity \citep{1996ASPC...93..174H}. The temperature of the ejected matter is held constant ($T \approx 10^{4}$ K) in time and radius during outburst \citep{Ribeiro}. The density gradient in these models usually assumes an inverse-square law, which corresponds to uniformly distributed mass as a function of expansion velocity \citep{Seaquist_Bode08}. A density gradient proportional to an inverse cube law produces similar results \citep{1980AJ.....85..283S}. In the case of inverse-square distribution, the ejecta density $\rho$ at an arbitrary radius $R$ (between $R_{\rm in}$ and $R_{\rm out}$) is given by
\begin{equation}
\rho(R,t) = \frac{1}{4\pi R^{2}}\times \frac{M_{\rm ej}}{R_{\rm{out}}(t)-R_{\rm in}(t)}\
\end{equation}
where $M_{\rm ej}$ is the total ejected mass \citep{Seaquist_Bode08}. Using this analytic form for the evolving density profile makes it fairly straightforward to calculate the radio flux density and opacity using basic numerical integration methods \citep{1977ApJ...217..781S,1979AJ.....84.1619H,1980AJ.....85..283S}.

\section{Observations and Data Reduction}

V351~Pup was observed by the pre-upgrade VLA at four frequency bands (C, X, U, and K bands), spanning 4.9 GHz--22.5 GHz following the 1991 nova outburst. Radio observations were obtained by R.\ Hjellming between the dates of 1992 March 7 and 1995 January 7 through the programs AH0390 and AH0492. The data were calibrated using the absolute flux calibrator 3C286 and the complex gain calibrator 0741-063. Data were obtained in standard continuum mode, with two intermediate frequencies (IFs) and 100 MHz bandwidth.

The raw data were accessed from the NRAO archive and were edited, calibrated, and imaged using standard routines in the Astronomical Image Processing System (AIPS; \citealt{2003ASSL..285..109G}). In every image, the flux density of V351~Pup was measured by fitting a Gaussian to the imaged source with the task \tt JMFIT \rm in AIPS. The resulting multi-frequency radio light curve for V351~Pup can be seen in Figure 1. For data sets obtained in the A configuration (the most extended VLA configuration), we recorded the integrated flux density of the Gaussian, while all other configurations held the width of the Gaussian fixed to the dimensions of the synthesized beam. Measurements are presented in Table 1.

\begin{figure}
\begin{center}
\includegraphics[height=2.75in,angle=0]{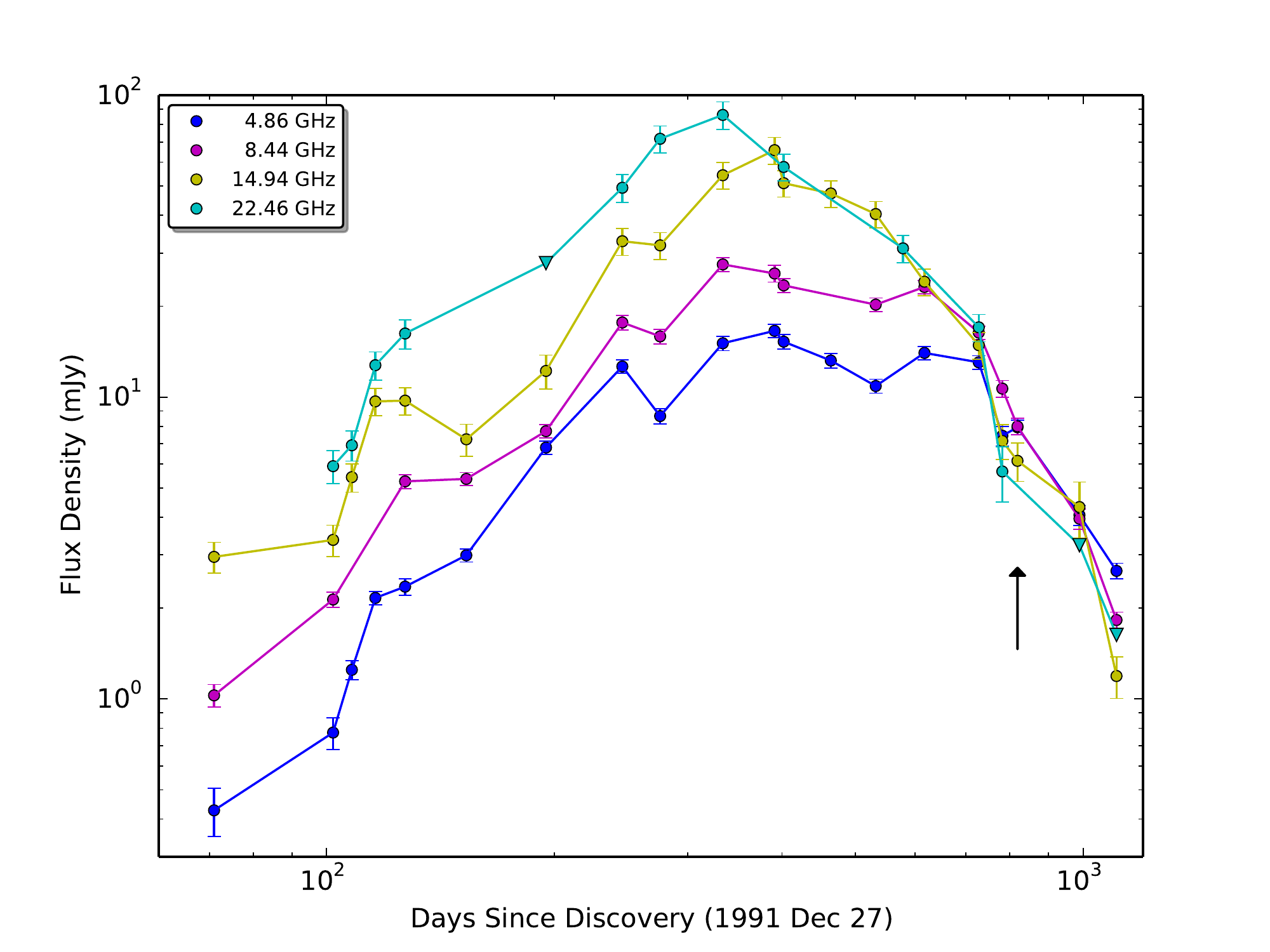}
\caption{Multi-frequency radio light curve for V351 Pup. Upper limits are plotted as downward-facing triangles (all seen at 22.5 GHz). The first epoch in which the source was resolved in the radio is indicated by the black arrow (see section \S 4.2 for further details).}
\end{center}
\end{figure}

V351~Pup has J2000 coordinates of R.A. = $08^{h}11^{m}38.38^{s}$, decl. = $-35^{\circ}07$'$30.4$", which puts 3C286 and 0741-063 at angular separations of $91^{\circ}$ and $140^{\circ}$ respectively. The large angular separation between V351~Pup and the complex gain calibrator had the potential to cause errors during calibration. The effects of these calibrations' errors are noted for
individual epochs in Table 1 with footnotes. We added 5\% systematic calibration errors to measurements at 4.9 and 8.4 GHz, and 10\% for 14.9 and 22.5 GHz.

An attempt was made to self-calibrate every epoch, but there was only sufficient signal-to-noise to successfully self-calibrate around the light curve peak (day 276 to day 728). Poor atmospheric conditions with the data sets on day 300 and the 22.5 GHz band on day 617 led to V351 Pup being fully decorrelated in the resulting images, and therefore no flux densities were recorded for that date.

3C286 was not observed as part of data sets on day 72, day 391, and at 22.5 GHz on day 102. However, the flux density of the phase calibrator 0741-063 appears stable with time (Table 1), so we used the flux density measured for 0741-063 on adjacent dates to set the absolute flux scale for these data sets.

R.\ Hjellming previously published results for radio detections on day 71. He recorded flux densities of 0.5, 1.2, and 3 mJy at 4.9, 8.4, and 14.9 GHz respectively \citep{1992IAUC.5473....3H}. We produced similar results for day 71 with flux densities of 0.43, 1.03, and 2.96 mJy corresponding to the same frequencies. The specific procedure for flux calibration used by R.\ Hjellming is unknown, but we suspect differences in our measurements to be caused by calibration without the flux calibrator 3C286. We assumed a constant flux density for the phase calibrator 0741-063 from the adjacent epoch on day 102.

V351~Pup was not significantly detected ($< 3 \sigma$) at 22.5 GHz on days 195, 989, and 1107. We suspect that poor weather conditions on day 195 led to low signal-to-noise. For the other two epochs, we suspect that the source had faded below the detection threshold. In these cases, the flux densities at the nova position were measured directly from the images. The rms noise levels for these sets were measured with \tt IMEAN \rm in AIPS. The upper limits plotted in Figure 1 and listed in Table 1 represent 3 $\sigma$ limits. Three times the noise was added to the flux density measured at the nova position, and that resulting calculation was plotted as the downward-facing triangle upper-limit.

Finally, the 22.5 GHz observation on day 819 and the 14.9 / 22.5 GHz data sets from day 875, when the VLA was in its extended A configuration, resulted in no recorded flux density due to V351~Pup being resolved out of the image. The A configuration of the VLA provides high resolution and is therefore beneficial for spatially resolved images. As discussed in \S 4.2, the observations from day 819 provided images at 8.4 and 14.9 GHz.

The radio light curve in Figure 1 takes approximately one year to reach maximum. As theoretically expected for thermal emission, higher frequencies peak earlier and at higher flux densities.
We note an early bump in the light curve, at 115 days after outburst and seen in multiple frequencies, which is followed by possible substructure at later times. Given the evidence for multiple ejecta events in other nova light curves \citep{1987A&A...183...38T, 2011ApJ...739L...6K, 2015arXiv151006751W, 2015arXiv150505879W}, these suggest that similar events may have occurred in V351~Pup. The large separation between the nova and the complex gain calibrator, however, makes us wary to interpret smaller-scale structures in the radio light curve in detail.

\LongTables
\begin{deluxetable*}{cccccccc}
\tablewidth{0pt}
\tabletypesize{\footnotesize}
\tablecaption{ \label{tab:swift}
Recorded Flux Densities for V351 Pup}
\tablehead{Observation & Time Since & Frequency & Frequency & Peak
Flux Density & Peak Flux Density & Array & Phase Calibrator \\
Date & Outburst (Days)\tablenotemark{1} & (GHz) & Band & (mJy) & Error
(mJy) & Configuration & Flux Density (mJy) }
\startdata
1992 Mar 7 & 71 & 4.9 & C & 0.43 \tablenotemark{2} & 0.08 & C & 2.79 \tablenotemark{2}
\\
1992 Mar 7 & 71 & 8.4 & X & 1.03 \tablenotemark{2} & 0.09 & C & 1.55 \tablenotemark{2}
\\
1992 Mar 7 & 71 & 14.9 & U & 2.96 \tablenotemark{2} & 0.35 & C & 0.83
\tablenotemark{2} \\
\\
1992 Apr 7 & 102 & 4.9 & C & 0.77 & 0.09 & C & 2.79
\\
1992 Apr 7 & 102 & 8.4 & X & 2.13 & 0.13 & C & 1.55
\\
1992 Apr 7 & 102 & 14.9 & U & 3.36 & 0.40 & C & 0.80
\\
1992 Apr 7 & 102 & 22.5 & K & 5.90 \tablenotemark{2} & 0.74 & C & 0.57 \tablenotemark{2} \\
\\
1992 Apr 13 & 108 & 4.9 & C & 1.25 & 0.09 & C & 2.80
\\
1992 Apr 13 & 108 & 14.9 & U & 5.43 & 0.59 & C & 0.81
\\
1992 Apr 13 & 108 & 22.5 & K & 6.92 & 0.79 & C & 0.53 \\
\\
1992 Apr 21 & 116 & 4.9 & C & 2.16 & 0.11 & C & 2.79
\\
1992 Apr 21 & 116 & 14.9 & U & 9.68 & 0.99 & C & 0.86
\\
1992 Apr 21 & 116 & 22.5 & K & 12.75 & 1.36 & C & 0.57 \\
\\
1992 May 2 & 127 & 4.9 & C & 2.35 & 0.15 & C & 2.77
\\
1992 May 2 &127 & 8.4 & X & 5.26 & 0.28 & C & 1.51
\\
1992 May 2 & 127 & 14.9 & U & 9.73 & 1.01 & C & 0.79
\\
1992 May 2 & 127 & 22.5 & K & 16.24 & 1.80 & C & 0.55\\
\\
1992 May 28 & 153 & 4.9 & C & 2.99 & 0.15 & CD & 2.80
\\
1992 May 28 & 153 & 8.4 & X & 5.36 & 0.27 & CD & 1.55
\\
1992 May 28 & 153 & 14.9 & U & 7.25 & 0.88 & CD & 0.83 \\
\\
1992 Jul 9 & 195 & 4.9 & C & 6.81 & 0.34 & D & 2.91
\\
1992 Jul 9 & 195 & 8.4 & X & 7.71 & 0.39 & D & 1.69
\\
1992 Jul 9 & 195 & 14.9 & U & 12.21 & 1.57 & D & 1.03
\\
1992 Jul 9 & 195 & 22.5 & K & $<$12.30 \tablenotemark{3} & 5.30 & D & 0.59 \\
\\
1992 Aug 29 & 246 & 4.9 & C & 12.63 & 0.64 & D & 2.79
\\
1992 Aug 29 & 246 & 8.4 & X & 17.66 & 1.00 & D & 1.57
\\
1992 Aug 29 & 246 & 14.9 & U & 32.83 & 3.38 & D & 0.82
\\
1992 Aug 29 & 246 & 22.5 & K & 49.36 & 5.24 & D & 0.57 \\
\\
1992 Sep 28 & 276 & 4.9 & C & 8.65 \tablenotemark{4} & 0.51 & D & 2.78
\\
1992 Sep 28 & 276 & 8.4 & X & 15.84 \tablenotemark{4} & 0.87 & D & 1.57
\\
1992 Sep 28 & 276 & 14.9 & U & 31.81 \tablenotemark{4} & 3.25 & D & 0.81
\\
1992 Sep 28 & 276 & 22.5 & K & 71.72 \tablenotemark{4} & 7.41 & D & 0.52 \\
\\
1992 Nov 25 & 334 & 4.9 & C & 15.07 \tablenotemark{4} & 0.82 & A & 2.81
\\
1992 Nov 25 & 334 & 8.4 & X & 27.51 \tablenotemark{4} & 1.45 & A & 1.53
\\
1992 Nov 25 & 334 & 14.9 & U & 54.31 \tablenotemark{4} & 5.49 & A & 0.78
\\
1992 Nov 25 & 334 & 22.5 & K & 86.02 \tablenotemark{4} & 9.01 & A & 0.52\\
\\
1993 Jan 21 & 391 & 4.9 & C & 16.58 \tablenotemark{2,4} & 0.85 & A &2.80 \tablenotemark{2}
\\
1993 Jan 21 & 391 & 8.4 & X & 25.67 \tablenotemark{2,4} & 1.64 & A & 1.53 \tablenotemark{2}
\\
1993 Jan 21 & 391 & 14.9 & U & 65.75 \tablenotemark{2,4} & 6.68 & A & 0.79 \tablenotemark{2} \\
\\
1993 Feb 1 & 402 & 4.9 & C & 15.26 \tablenotemark{4} & 0.84 & AB & 2.79
\\
1993 Feb 1 & 402 & 8.4 & X & 23.44 \tablenotemark{4} & 1.26 & AB & 1.53
\\
1993 Feb 1 & 402 & 14.9 & U & 51.13 \tablenotemark{4} & 5.13 & AB & 0.77
\\
1993 Feb 1 & 402 & 22.5 & K & 57.84 \tablenotemark{4} & 5.83 & AB &
0.46 \\
\\
1993 Apr 4 & 464 & 4.9 & C & 13.21 \tablenotemark{4} & 0.74 & B & 2.79
\\
1993 Apr 4 & 464 & 14.9 & U & 49.00 \tablenotemark{4} & 4.75 & B & 0.82 \\
\\
1993 Jun 11 & 532 & 4.9 & C & 10.88 \tablenotemark{4} & 0.58 & C & 2.81
\\
1993 Jun 11 & 532 & 8.4 & X & 20.24 \tablenotemark{4} & 1.05 & C &1.58
\\
1993 Jun 11 & 532 & 14.9 & U & 40.39 \tablenotemark{4} & 4.05 & C &0.88 \\
\\
1993 Jul 27 & 578 & 22.5 & K & 31.09 \tablenotemark{4} & 3.19 & C &
0.48\\
\\
1993 Sep 4 & 617 & 4.9 & C & 14.00 \tablenotemark{4} & 0.71 & CD & 2.80
\\
1993 Sep 4 & 617 & 8.4 & X & 23.16 \tablenotemark{4} & 1.16 & CD & 1.56
\\
1993 Sep 4 & 617 & 14.9 & U & 24.11 \tablenotemark{4} & 2.43 & CD & 0.80 \\
\\
1993 Dec 24 & 728 & 4.9 & C & 13.04 \tablenotemark{4} & 0.67 & D
& 2.78
\\
1993 Dec 24 & 728 & 8.4 & X & 16.35 \tablenotemark{4} & 0.82 & D & 1.56
\\
1993 Dec 24 & 728 & 14.9 & U & 14.86 \tablenotemark{4} & 1.52 & D &0.80
\\
1993 Dec 24 & 728 & 22.5 & K & 17.02 \tablenotemark{4} & 1.76 & D
&0.49 \\
\\
1994 Feb 16 & 782 & 4.9 & C & 7.45 & 0.55 & D & 2.80
\\
1994 Feb 16 & 782 & 8.4 & X & 10.67 & 0.69 & D & 1.56
\\
1994 Feb 16 & 782 & 14.9 & U & 7.15 & 0.94 & D & 0.77
\\
1994 Feb 16 & 782 & 22.5 & K & 5.67 & 1.18 & D & 0.51 \\
\\
1994 Mar 25 & 819 & 4.9 & C & 7.94 & 0.45 & A & 2.75
\\
1994 Mar 25 & 819 & 8.4 & X & 8.00 & 0.50 & A & 1.51
\\
1994 Mar 25 & 819 & 14.9 & U & 6.15 & 0.89 & A & 0.75
\\
1994 Mar 25 & 819 & 22.5 & K & \nodata \tablenotemark{5}& \nodata & A
& 0.47 \\
\\
1994 May 20 & 875 & 14.9 & U &\nodata \tablenotemark{5} & \nodata &
AB & 0.85
\\
1994 May 20 & 875 & 22.5 & K & \nodata \tablenotemark{5} & \nodata &
AB & 0.64 \\
\\
1994 Sep 11 & 989 & 4.9 & C & 4.07 & 0.31 & B & 2.80
\\
1994 Sep 11 & 989 & 8.4 & X & 3.95 & 0.31 & B & 1.55
\\
1994 Sep 11 & 989 & 14.9 & U & 4.32 & 0.91 & B & 0.81
\\
1994 Sep 11 & 989 & 22.5 & K & $<$0.56 \tablenotemark{3} & 0.89 & B
& 0.52 \\
\\
1995 Jan 7 & 1107 & 4.9 & C & 2.66 & 0.12 & CD & 2.79
\\
1995 Jan 7 & 1107 & 8.4 & X & 1.83 & 0.11 & CD & 1.55
\\
1995 Jan 7 & 1107 & 14.9 & U & 1.19 & 0.19 & CD & 0.80
\\
1995 Jan 7 & 1107 & 22.5 & K & $<$0.70 \tablenotemark{3} & 0.32 & CD & 0.50 \\
\enddata
\tablenotetext{1}{Nova outburst was discovered on 1991 December 27, which we take to be $t_0$ \citep{1992IAUC.5422....1C}.}
\tablenotetext{2}{This observation was calibrated without the flux calibrator 3C286, assuming the listed flux density for the phase calibrator, derived using the flux measured from adjacent epochs.}
\tablenotetext{3}{ The source did not significantly exceed the noise. The quoted value is a 3$\sigma$ upper limit.}
\tablenotetext{4}{Data were partially decorrelated and, as a result, were self-calibrated to recover the full flux density.}
\tablenotetext{5}{The source was resolved out by the extended array configuration.}
\end{deluxetable*}

\vspace{0.5cm}

Figure 2 shows the behavior of the radio spectral energy distribution of V351~Pup as the nova shell expands. As previously stated, the
spectral index is expected to be high when completely optically thick ($\alpha \approx 2.0$) and low when optically thin ($\alpha \approx -0.1$). V351~Pup's spectral index peaks at $\alpha=1.31 \pm 0.07$ on day 276 days after outburst,
and remains roughly constant between $\alpha = 0.8-1.2$ until day $\sim$400. A true spectral index of $\alpha=2.0$ is not often observed in novae, and a spectral index of $\alpha \approx 1.3$ is more commonly observed during the early evolution of novae
\citep{2014Natur.514..339C,2014ApJ...785...78N}. The underlying cause of this discrepancy from theoretical expectations is not yet understood.

The data taken before day 402 can be reasonably fit with a single power law. On day 402, V351~Pup begins to show evidence of turnover at high frequencies, as the 22.5 GHz band falls below the power-law fit to the other frequencies. This observation demonstrates that by day 402, the highest frequency photosphere has started to recede back through the nova shell. Unfortunately, there is no clear turnover on day 532 because the VLA did not observe V351 Pup at 22.5 GHz. By day 728, the radio photosphere has become optically thin at all frequencies, yielding a roughly flat spectral index of $\alpha=0.17 \pm 0.07$.

\begin{figure*}[!]
\begin{center}
\includegraphics[height=5.4in,angle=0]{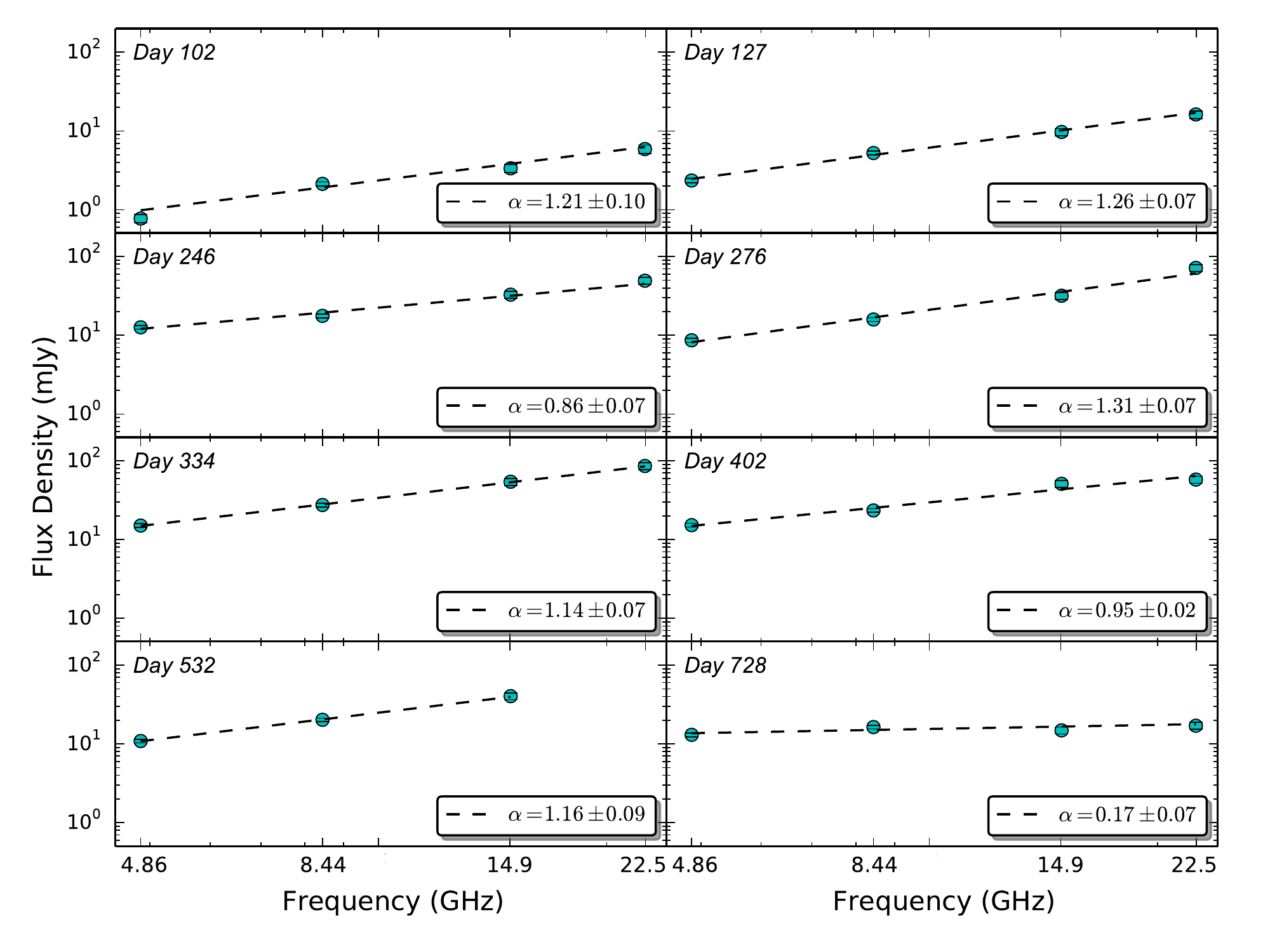}
\caption{Plots of V351~Pup's radio spectrum at eight different epochs (teal data points). Each epoch was fit with a single power law (black dashed line). Best-fit spectral indices and uncertainties are given in the bottom-right of each panel.}
\end{center}
\end{figure*}

\section{Imaging of V351 Pup}

\subsection{\textit{Optical Imaging}}

Pictured in Figure 3 is an optical narrow-band image of V351~Pup taken with the \emph{HST} on 1998 February 14 (day 2,240). The image was taken on the PC chip of the WFPC2 camera with the $F656N$ filter and an exposure time of 320 s. The $F656N$ filter covers H$\alpha$ and the [\ion{N}{2}] 6548 \AA\ and 6584 \AA\ lines. The shell of ejected matter can be seen as a circularly symmetric ring of material surrounding the WD star. We note that this image was considered in \citet{2000AJ....120.2007D} for distance calculations, but was not pictured therein.

To measure the diameter of the nova shell, we took an image slice across the shell's center (intersecting the white-dwarf point source) of 1 pixel width. Along this image slice, the shell was visible as a flux peak on either side of the point source. We fit Gaussians to these maxima using the \verb|curve_fit| function in the Python scipy package. Two different image slices were taken, oriented orthogonal to one another, to yield four total measurements of the shell's gaussian profile. The four total measurements of the gaussian width were averaged, yielding a full width at half maximum (FWHM) for the shell of $0.38 \pm 0.04^{\prime\prime}$. The diameter was calculated by adding two times the standard deviation to the difference in mean values along each image slice. We measured the diameter of the nova shell to be $1.2 \pm 0.1^{\prime\prime}$, as indicated by the red circle in Figure 3.

\begin{figure}
\begin{center}
\includegraphics[width=3.4in,angle=0]{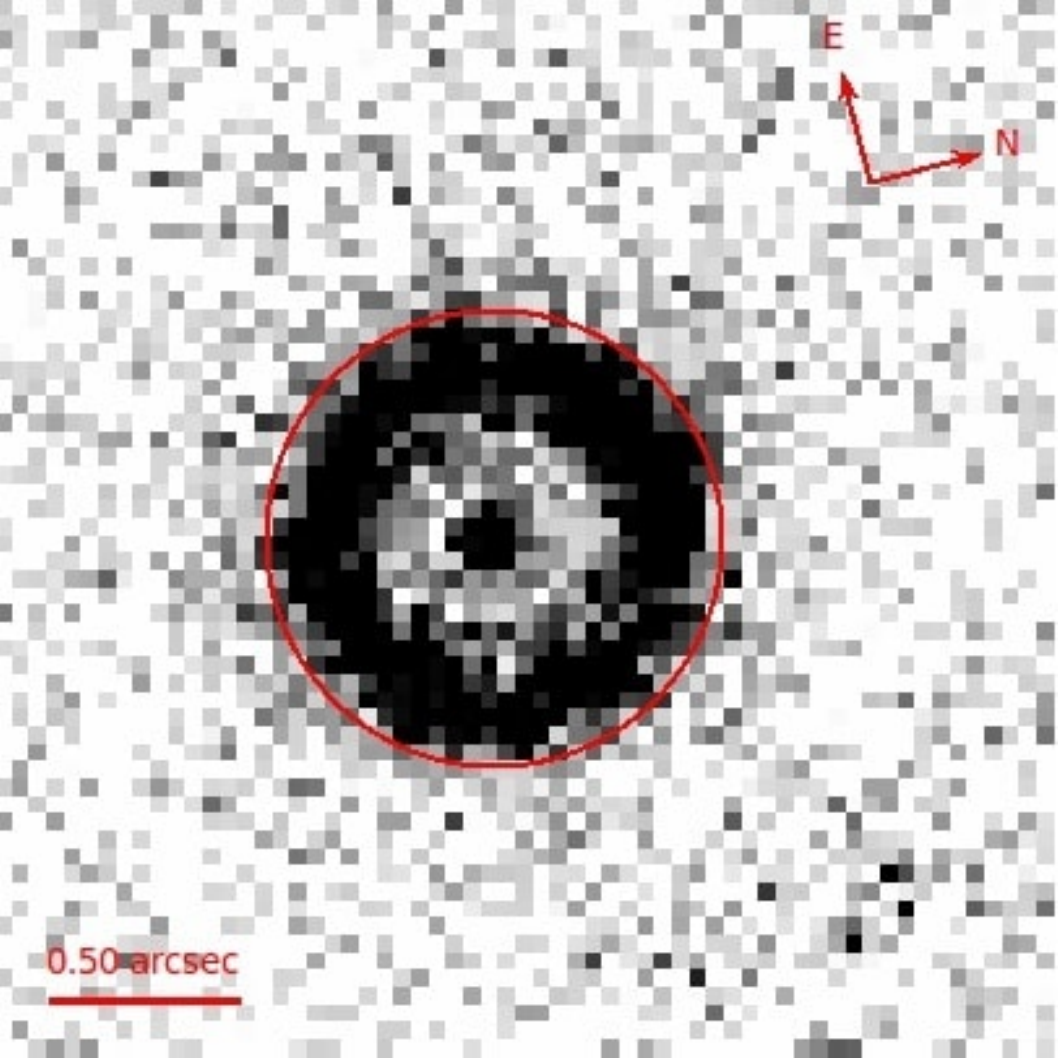}
\caption{$F658N$ image of V351~Pup taken with \emph{HST} on 1998 February 14. The extent of V351 Pup's circularly symmetric ejected shell is indicated by the red circle. The diameter of the circular ring is $1.2 \pm 0.1^{\prime\prime}$. The field of view is $2.7^{\prime\prime} \times 2.3^{\prime\prime}$. }
\label{light_curve}
\end{center}
\end{figure}

\subsection{\textit{Radio Imaging}}

VLA A-configuration imaging collected 819 days after outburst displays evidence that V351~Pup has become resolved at 8.4 and 14.9 GHz, while V351~Pup was not clearly resolved at 4.9 GHz, and its flux was resolved out at 22.5 GHz. Figure 4 displays radio images produced in AIPS for this epoch; to the left is 14.9 GHz and to the right is 8.4 GHz. The top-right in each panel shows a white ellipse representing the synthesized beam. The dimensions of the synthesized beam at 14.9 GHz are $0.40^{\prime\prime} \times 0.16 ^{\prime\prime}$ (FWHM) with a position angle of $-5.6^{\circ}$. At 8.4 GHz, the synthesized beam measures $0.55^{\prime\prime} \times 0.20^{\prime\prime}$ with a position angle of $-2.6^{\circ}$. V351~Pup is clearly less resolved at 8.4 GHz, compared to 14.9 GHz.

To determine the size and level of asymmetry of the radio image, we modeled the calibrated $u-v$ data corresponding to the images in Figure 4 with Difmap \citep{1994BAAS...26..987S}. Both circular and elliptical 2D Gaussian components were fit to the 8.4 and 14.9 GHz data sets; we find that both models yield very similar residuals. Results can be seen in Table 2. While the elliptical models contain slightly more flux and have slightly lower $\chi^2$ values than the circular model, the difference is not significant. We conclude that the radio images of V351 Pup are consistent with a circular geometry of diameter $0.47^{\prime\prime}$, observed 819 days after nova explosion.

\begin{table}[hb]
\caption{Residuals from modeling data sets from day 819 after outburst with Difmap.}
\centering 

\setlength{\tabcolsep}{3.1pt}
\begin{tabular}{l r r l r r} 
\hline\hline 
& Circular & & & Elliptical & \\ [0.5 ex] 
\hline 
Frequency & 8.4 & 14.9 & Frequency & 8.4& 14.9 \\
(GHz)& & &(GHz)& & \\
\\
Integrated & & & Integrated & \\
Flux Density &8.25 &6.14 & Flux Density &8.29 &6.23 \\
(mJy)& & &(mJy) & & \\
\\
& & & Major Axis & 0.57& 0.65\\
& & & FWHM ($^{\prime\prime}$) & & \\
\\
FWHM ($^{\prime\prime}$)& 0.46& 0.48& Minor Axis & 0.34& 0.36\\
& & & FWHM ($^{\prime\prime}$) & & \\
\\
& & & Position Angle &48.9 &27.8 \\
& & &($^{\circ}$) & & \\
\\
Reduced $\chi^2 $ &1.223 &1.084 & Reduced $\chi^2 $ &1.222 &1.084 \\ [0.1 ex]
\hline \\
\end{tabular}
\label{table:nonlin} 
\end{table}

The Hubble flow model (and most published modeling of novae at radio bands) assumes spherical geometry. \citet{Ribeiro} showed that departures from spherical symmetry (i.e., bipolar ejecta morphologies) can significantly affect the radio light curve. They demonstrate that a faulty assumption of spherical symmetry - when the ejecta are instead bipolar - can lead to an overestimate of the ejecta mass by a factor of $\sim$2. Such complications do not appear to be relevant for V351~Pup, however, as shown by both the radio and \emph{HST} imaging. V351~Pup shows no significant departures from spherical symmetry.


\subsection{\textit{Distance via Expansion Parallax}}

Spherical symmetry in V351~Pup also enables a relatively simple derivation of the distance using the expansion parallax. The expansion parallax technique uses spatially resolved imaging of an expanding body to determine the rate of angular expansion (i.e., arcsec~yr$^{-1}$ in the plane of the sky). Doppler shift measurements of radial velocity are used in conjunction to yield a physical expansion rate (km~s$^{-1}$). A comparison of the angular and physical expansion rates yield a distance to the expanding nova \citep[e.g.,][]{1987A&A...183...38T, 2000AJ....120.2007D, 2000MNRAS.318.1086E}. Non-spherical geometries complicate the comparison of radial and transverse velocity, leading to more involved expansion parallax calculations \citep{2000PASP..112..614W, 2009ApJ...706..738W, 2015ApJ...805..136L}. However, the aforementioned spherical symmetry allows us to assume that the transverse and radial velocity are equal.

V351~Pup has become optically thin at both 14.9 GHz and 8.4 GHz by day 819 (the time of our imaging), as can be seen by the flat radio spectrum measured on day 728 (Figure 2). Therefore, caution must be exhibited in using these observations to find a distance using the expansion parallax, because the outer, less dense ejecta (corresponding to the fastest velocities) could be faint and below the sensitivity limit of our imaging. However, these concerns are assuaged by the fact that very similar diameters are measured at both 14.9 and 8.4 GHz, despite rather different sensitivities and resolutions. We therefore conclude that we are sensitive to the outer edges of the nova ejecta, and their diameter is $0.47^{\prime\prime}$.

The radial expansion rate for the radio images was therefore found to be 0.105 arcsec yr$^{-1}$. The expansion rate for the H$\alpha$+[\ion{N}{2}] image was measured to be 0.097 arcsec yr$^{-1}$. Although we expect, to first order, that the H$\alpha$+[\ion{N}{2}] and radio imaging trace the same warm ionized gas, the dramatically different observing epochs and sensitivity might allow one to be more sensitive to the outermost, low-density ejecta than the other. It is therefore remarkable that the radio and \emph{HST} expansion rates agree so
well with one another. For the following expansion parallax calculation, we take the radio-determined expansion rate, as it was determined earlier and represents faster expansion, and therefore more likely traces the outermost ejecta.

These angular expansion rates should be compared to the maximum ejecta velocities, determined from spectroscopy, to determine an expansion parallax distance. We take the maximum velocity of the expanding shell to be 2500 km s$^{-1}$, as measured from the full width at zero intensity (FWZI) of the Balmer lines (\citealt{1992IAUC.5428....2S,1994ApJS...90..297W}). Faster expansion velocities were measured for the \ion{Mg}{2} line, but the Balmer lines are most directly comparable to the thermal free-free radio emission. Using Equation 1 from \citet{2015ApJ...805..136L}, we therefore estimate a distance to V351~Pup of 5.0 kpc.
By far the largest uncertainty in this calculation is the physical velocity. \citet{2000AJ....120.2007D} used the \emph{HST} image shown in \S 4.1 and assumed ``typical" (i.e., mean) ejecta velocities of 1200--2000 km s$^{-1}$, based off the work of \citet{1995ASSL..205..303P} and \citet{1992IAUC.5428....2S}. While they calculated a very similar angular expansion velocity as we find here (0.100 arcsec yr$^{-1}$), they derived a distance in the range of 2.7--4.0 kpc. We estimate a 30\% error in the physical expansion velocity corresponding to our radio and \emph{HST} images (i.e., 2500 $\pm$ 750 km s$^{-1}$), therefore implying a distance and uncertainty of $5.0 \pm 1.5$ kpc.

\citet{1996MNRAS.279..280S} published a few additional distance estimates based on independent techniques. They derived a distance from ultraviolet data, assuming V~351 Pup is observed during a phase of constant bolometric luminosity near Eddington; they estimate 2.4--3.0 kpc. This estimate falls $\sim$1.6$\sigma$ from our distance determination, and would be very susceptible to foreground reddening. \citet{1996MNRAS.279..280S} also consider distance estimates based off the maximum magnitude--rate of decline relationship, but we neglect those here based on the work of \citet{2011ApJ...735...94K} and \citet{2012ApJ...752..133C}, which brings the validity of the relationship into question.

\begin{figure}[!]
\begin{center}
\includegraphics[height=2.1in,angle=0]{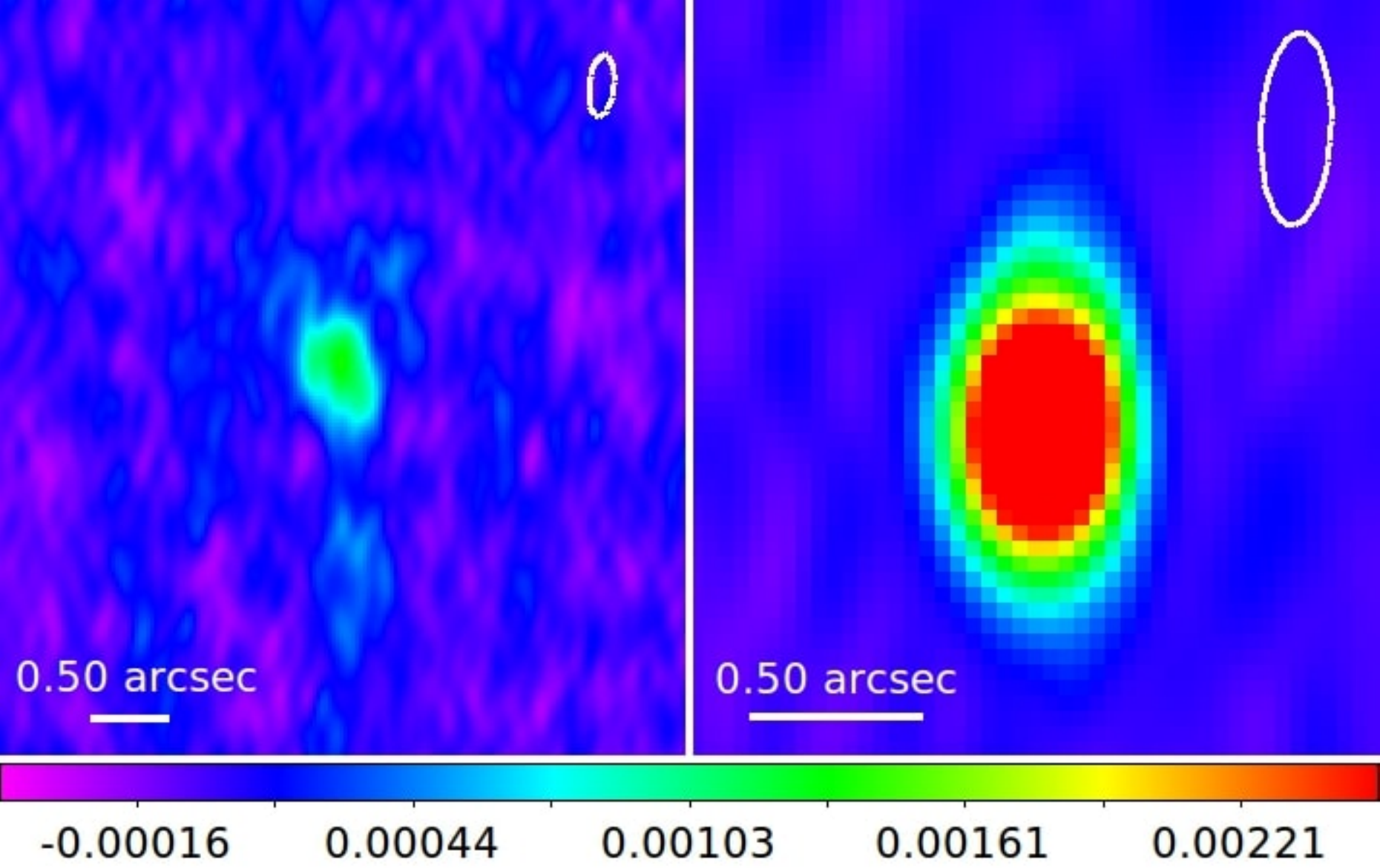}
\caption{Radio imaging taken 819 days after outburst (1994 March 25) at two different frequencies: 14.9 GHz (left panel) and 8.4 GHz (right panel). The synthesized beam is pictured as a white ellipse in the top right in each panel. The field of view for the 14.9 GHz image is $4.6^{\prime\prime} \times 4.8^{\prime\prime}$, and the field of view for the 8.4 GHz image is $2.0^{\prime\prime} \times 2.2^{\prime\prime}$. The flux scale was held constant between the two images and is given in units of jansky.}
\label{light_curve}
\end{center}
\end{figure}

\begin{figure*}
\begin{center}
\includegraphics[height=5.3in,angle=0]{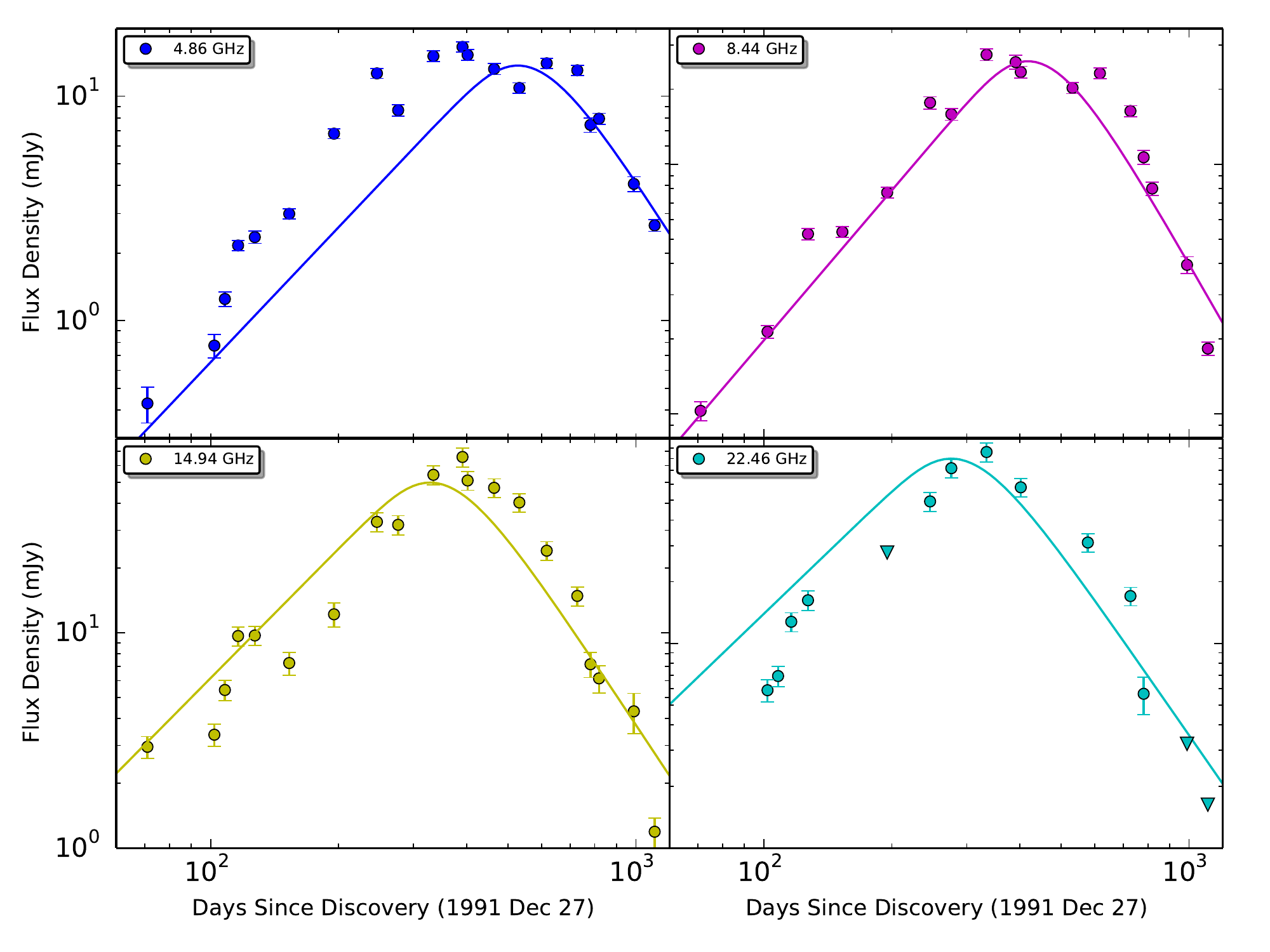}
\caption{Individual frequency radio light curves for V351 Pup with minimization of $\chi^2$ between our data and the Hubble flow model in linear space plotted as a solid line. The 3$\sigma$ upper limits plotted for the K-band frequency are indicated by the downward-facing triangles, but are not used in the fits.}
\label{light_curve}
\end{center}
\end{figure*}

\section{Light-curve Modeling}

The standard Hubble flow model can be applied to our light curve in
order to estimate the ejecta mass for V351 Pup.
The multi-frequency light curves for V351 Pup were fit with the
standard Hubble flow model via $\chi^2$ minimization, fitting $>$3$\sigma$ detections at all four
frequencies and all times simultaneously.
The model fit assumed that the instant of mass ejection was simultaneous with the optical discovery date on 1991 December 27 \citep{1992IAUC.5422....1C}.



We incorporate within our model physical parameters that have previously been derived in the literature.
The electron temperatures were derived by \citet{1996MNRAS.279..280S} from the flux ratios of optical and ultraviolet emission lines for nitrogen \citep{1981MNRAS.197..107S,1981MNRAS.197....1W,1996MNRAS.279..280S}. Measurements were taken between 136 and 484 days after outburst and show no evidence for a significant temperature change over time; they are always consistent with $\sim 10^4$ K, which we take as the electron temperature in our model.

\citet{1996MNRAS.279..280S} calculated volume filling factors that are on the order of $\epsilon = 10^{-5}$, by measuring the electron density using [\ion{O}{3}] line ratios (assuming the electron temperature as described above) and comparing them with estimates of the emission measure ($\propto$ density$^2$) determined from hydrogen recombination lines.
Filling factors were calculated for dates between 370 and 484 days after outburst and remain consistent around 10$^{-5}$.

The derived mass of the ejecta depends on the filling factor to the power of 0.5 \citep{2005MNRAS.362..469H} and on the distance to the power of 2.5 \citep{1979AJ.....84.1619H}. Based on our estimates in \S 4.3, we assume a distance of $5.0 \pm 1.5$ kpc (a $\sim$30\% error). In addition, the filling factor is known to be good to one order of magnitude \citep{1996MNRAS.279..280S}. Given the large uncertainty in our distance and the filling factor, we varied these parameters to find a range for our ejecta mass. To find an average value of the mass ejecta, we assumed a distance of 5 kpc and a filling factor of $10^{-5}$. For the minimum mass ejecta, we assumed a distance of 3.5 kpc and a filling factor of $10^{-6}$; and for the maximum mass ejecta we assumed a distance of 6.5 kpc and a filling factor of $10^{-4}$.

Model fits to each frequency are seen in Figure 5.
Our model fits for the expanding isothermal shell of (free-free) emitting gas fits three primary components of the light curve. These components include the total brightness, the amount of time it takes for the gas to become optically thin, and the speed with which it becomes optically thin.
Minimization of $\chi^2$ between our data and the Hubble flow model results in the model fit seen as the solid line. The maximum velocity of the ejecta from this model fit was calculated to be 3000
km~s$^{-1}$ and the ratio of velocities of the inner and outer end of the
ejected material, $\zeta$, was calculated to be 0.74. The mass of
ejecta was calculated to be $log_{10}(M_{ej})={-5.2} \pm {0.7}$ M$_{\odot}$. As pointed out above, the mass of the shell is proportional to distance$^{5/2}$, while the maximum velocity depends linearly on distance. If V351~Pup is at 2.5 kpc rather than 5 kpc, our estimates of ejecta mass would decrease to 18\% of those listed above (i.e., $[0.9-1.3] \times 10^{-6}$ M$_{\odot}$).

Our model fit, while not formally acceptable, was optimized at a
$\chi^2/\nu$ of $2074.8/65$ or a reduced $\chi^2$ of 31.9. Model fits
to multiple-frequency light curves essentially always return reduced
$\chi^2$ values greater than 1, illustrating that the simple Hubble
flow model is not completely sufficient to describe our data \citep[e.g.,][]{2014ApJ...785...78N,2016MNRAS.457..887W,2017arXiv170103094F}. Still,
it is outside the scope of this work to develop a model that can
completely describe the radio light curve, which would need to be more complex.

In a similar manner to how we calculated the diameter of the nova shell in \S 4.1, we were able to measure $\zeta$ at the time of our \emph{HST} image (day 2,240). $R_{\rm out}$ was calculated by adding two times the standard deviation to the difference in mean values along each image slice, and $R_{\rm in}$ was calculated by subtracting two times the standard deviation to the difference in mean values along each image slice. We measured our ratio of $R_{\rm in}$ to $R_{\rm out}$, also known as $\zeta$, to be $0.45 \pm 0.03$.

\section{Discussion}

Plotted in Figure 6 is nova ejecta mass against $t_{3}$, the time for
the optical light curve to decline by three magnitudes. The $t_{3}$ parameter is
often considered a proxy for WD mass
\citep{2005ApJ...623..398Y}, though it is subject to
poorly understood mass-loss physics in novae \citep{2013ApJ...777..136W}. Observed here is a factor of $>$10 discrepancy between observational and theoretical ejecta mass estimates \citep{1998MNRAS.296..502S}. Theoretical
predictions from \citet{2005ApJ...623..398Y} are shown as diamonds, and are
color-coded by the mass of the WD that hosts the explosion
(red $= 0.4$ M$_{\odot}$, blue $= 0.65$ M$_{\odot}$, purple $= 1.0$
M$_{\odot}$, cyan $= 1.25$ M$_{\odot}$, yellow $= 1.4$
M$_{\odot}$). Past observational estimates are plotted as light gray circles, with estimates of ejecta mass coming from radio data as compiled by \citet{Seaquist_Bode08} and T. L. Johnson et al.\ (2017, in preparation) and presented by \citet{2012BASI...40..293R}. All of these past observational estimates plotted assume a uniform filling factor. Recent eNova publications from \citet{2014ApJ...785...78N}, \citet{2016MNRAS.457..887W}, and \citet{2017arXiv170103094F} are plotted as well. \citet{2016MNRAS.457..887W} assumed a uniform filling factor, and therefore V1723 Aql is plotted as an upper limit. The analysis from \citet{2014ApJ...785...78N} shows that the ejecta mass from the recurrent nova T Pyx, plotted as a charcoal gray circle, varies between (1-30)$\times 10^{-5}$ M$_{\odot}$ when accounting for a plausible range of volume ﬁlling factors. \citet{2017arXiv170103094F} assumed a filling factor of $\epsilon = 5.2 \pm 2.2 \times 10^{-2}$ to derive an ejecta mass of $2.7 \pm 0.9 \times 10^{-4} M_{\odot}$ for V1324 Sco. Our estimate for ejecta mass place V351 Pup's WD mass in the range of 1.25 M$_{\odot}$, which is consistent with spectroscopic evidence for an oxygen-neon white dwarf \citep{1996MNRAS.279..280S}.

\begin{figure*}
\begin{center}
\includegraphics[height=5.3in,angle=0]{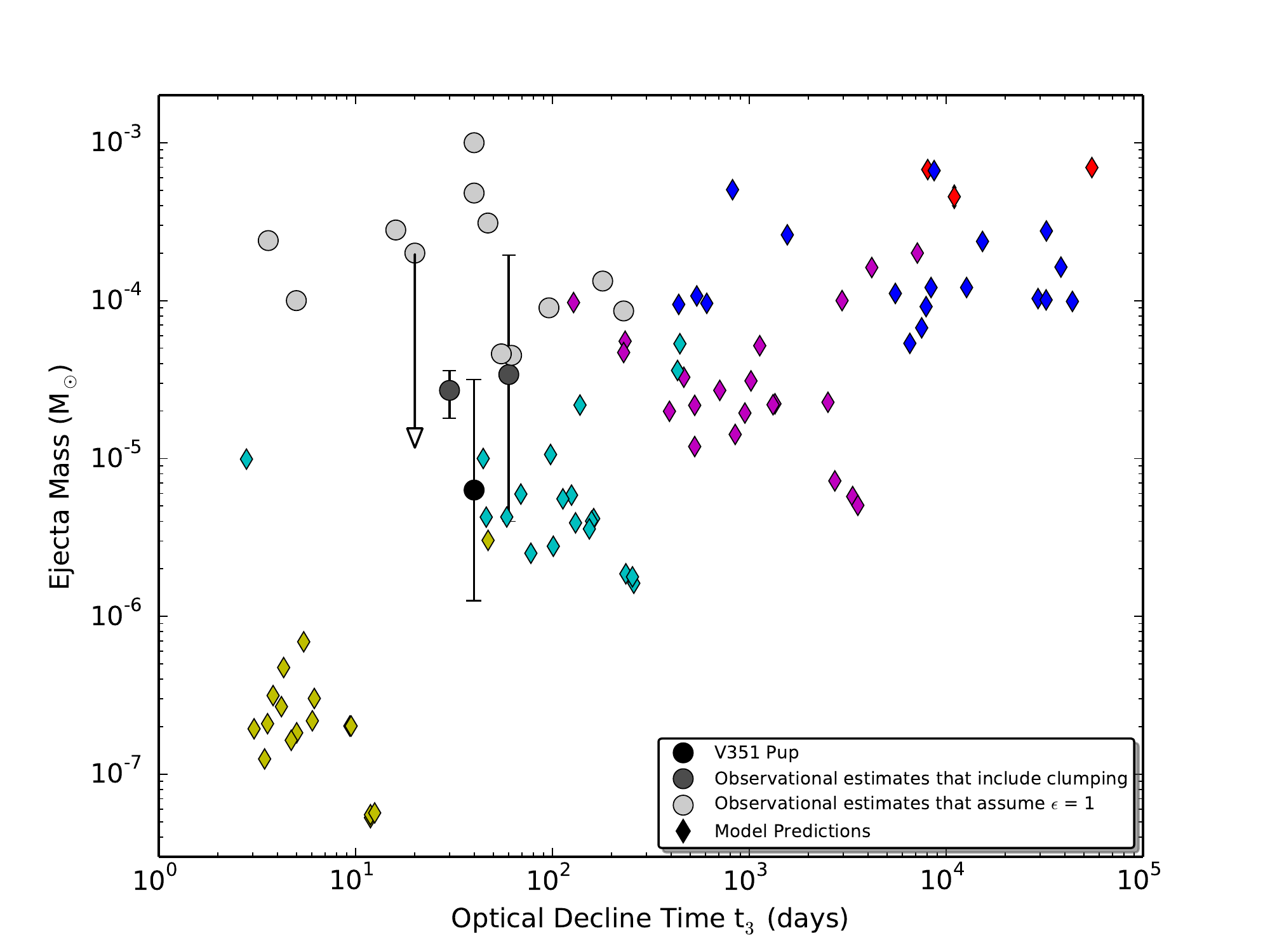}
\caption{Nova ejecta mass plotted against the time for the optical light curve
to decline by three magnitudes. Theoretical predictions from Yaron
et al. (2005) are shown as diamonds, and are color-coded by the mass
of the WD that hosts the explosion (red $= 0.4$
M$_{\odot}$, blue $= 0.65$ M$_{\odot}$, purple $= 1.0$ M$_{\odot}$,
cyan $= 1.25$ M$_{\odot}$, yellow $= 1.4$
M$_{\odot}$). Past observational estimates from radio data are plotted as
light gray circles, as compiled by Seaquist $\&$ Bode (2008) and T. L. Johnson et
al. (2017, in preparation). Recent eNova publications for the novae T Pyx and V1324 Sco are plotted as charcoal gray circles \citep{2014ApJ...785...78N,2017arXiv170103094F}
The recent eNova publication for V1723 Aql assumed a uniform filling factor, but discussed the notion of using a filling factor that incorporated clumping, so it is plotted as an upper-limit light gray circle \citep{2016MNRAS.457..887W}. Previously, Hjellming measured the mass of V351
Pup to be $1.01 \times 10^{-3}$ M$_{\odot}$ under the assumption of
uniform density ejecta (shown as one of the light gray circles). The best-fit estimate from our model
fitting that incorporated clumping places V351 Pup at a mass of $log_{10}(M_{ej})={-5.2} \pm {0.7}$ M$_{\odot}$ (shown as the black circle).}
\end{center}
\end{figure*}

Previously, R. Hjellming measured the mass of V351 Pup to be $1.01
\times 10^{-3}$ M$_{\odot}$ under the assumption of uniform density
ejecta. R. Hjellming assumed a maximum velocity of the expanding
shell to be 2500 km s$^{-1}$ and a distance of 4.1 kpc
\citep{1996ASPC...93..174H}. Holding Hjellming's parameters fixed we
find that this model produces a significantly worse fit than the model
we present here. This can be quantified by
$\chi^2$ statistics of how far his model deviants from the data. These parameters fit the data poorly with a
$\chi^2/\nu$ of 6541.0/65 or a reduced $\chi^2$ of 100.6. This value of
$\chi^2$ is roughly a factor of 3.2 worse than either of the best-fit
models we were able to produce. The parameters R. Hjellming used produced a
model that was not reaching high enough peak flux levels and declined
too slowly with respect to our light curve.

The estimate from our model fitting that incorporated clumping and errors in distance places V351 Pup's ejecta at a mass
of $log_{10}(M_{ej})={-5.2} \pm {0.7}$ M$_{\odot}$. It is worth noting that a non-unity filling factor does
not change the shape of the radio light curve, it only changes the
ejecta mass that corresponds to a particular light curve ($M_{ej}
\propto \epsilon^\frac{1}{2}]$) \citep{2014ApJ...785...78N}. If we assumed a
filling factor of 1, the best-fit ejecta mass described above would
become $1.48 \times 10^{-3}$, which is on the same order of magnitude as the ejecta mass derived
by R. Hjellming.
With clumping, the ejecta of V351~Pup have gone from being one
of the most massive ejections reported in the literature to several orders
of magnitude smaller, and are now consistent with nova model
predictions (Figure 6). Incorporating non-uniform filling factors into other
observed radio light curves has the potential to fix the long debated
discrepancy between theoretical and observed ejecta mass estimates in novae.


\section{Conclusions}
We observed V351 Pup to be a moderately well-behaved classical nova at radio
wavelengths, well described by thermal free-free emission.
Our model fit of expanding thermal ejecta held the
assumption of spherical symmetry, and an analysis of both radio and
H$\alpha$+[\ion{N}{2}] images of V351 Pup revealed no aspherical structure.
The mass of the ejecta was estimated to be $log_{10}(M_{ej})={-5.2} \pm {0.7}$ M$_{\odot}$ .
Previously, V351 Pup was published as one of the most massive ejections, but the incorporation of a
filling factor of $10^{-5}$ into our model now places V351 Pup on the low end of expectations for ejecta mass from classical novae, and brings its observational properties in line with theoretical predictions.

\acknowledgements
C.\ W.\ would like to thank the Department of Physics and Astronomy at Michigan State University for the Lawrence W.\ Hantel Endowed Fellowship Fund, in Memory of Professor Donald J.\ Montgomery; the Lyman Briggs College for the Undergraduate Research Scholarship; and the Michigan Space Grant Consortium Fellowship Program. Special thanks to Koji Mukai and Tommy Nelson for helpful insights. L.\ C.\ and J.\ D.\ L.\ are supported by NASA Fermi Guest Investigator grant NNH13ZDA001N-FERMI.

The National Radio Astronomy Observatory is a facility of the National Science Foundation operated under cooperative agreement by Associated Universities, Inc. This paper made use of observations made with the NASA/ESA Hubble Space Telescope, obtained from the data archive at the Space Telescope Science Institute. STScI is operated by the Association of Universities for Research in Astronomy, Inc.\ under NASA contract NAS 5-26555.

\bibliography{v351puppis.astroph}

\begin{thebibliography}{}
\expandafter\ifx\csname natexlab\endcsname\relax\def\natexlab#1{#1}\fi

\bibitem[{{Camilleri} {et~al.}(1992){Camilleri}, {McNaught}, {Gilmore}, \&
  {Kilmartin}}]{1992IAUC.5422....1C}
{Camilleri}, P., {McNaught}, R.~H., {Gilmore}, A.~C., \& {Kilmartin}, P.~M.
  1992, \iaucirc, 5422, 1

\bibitem[{{Cao} {et~al.}(2012){Cao}, {Kasliwal}, {Neill}, {Kulkarni}, {Lou},
  {Ben-Ami}, {Bloom}, {Cenko}, {Law}, {Nugent}, {Ofek}, {Poznanski}, \&
  {Quimby}}]{2012ApJ...752..133C}
{Cao}, Y., {Kasliwal}, M.~M., {Neill}, J.~D., {et~al.} 2012, \apj, 752, 133

\bibitem[{{Chomiuk} {et~al.}(2014){Chomiuk}, {Linford}, {Yang}, {O'Brien},
  {Paragi}, {Mioduszewski}, {Beswick}, {Cheung}, {Mukai}, {Nelson}, {Ribeiro},
  {Rupen}, {Sokoloski}, {Weston}, {Zheng}, {Bode}, {Eyres}, {Roy}, \&
  {Taylor}}]{2014Natur.514..339C}
{Chomiuk}, L., {Linford}, J.~D., {Yang}, J., {et~al.} 2014, \nat, 514, 339

\bibitem[{{Cunningham} {et~al.}(2015){Cunningham}, {Wolf}, \&
  {Bildsten}}]{2015ApJ...803...76C}
{Cunningham}, T., {Wolf}, W.~M., \& {Bildsten}, L. 2015, \apj, 803, 76

\bibitem[{{Della Valle} {et~al.}(1992){Della Valle}, {Reinsch}, {Thomas}, \&
  {Rampazzo}}]{1992IAUC.5427....1D}
{Della Valle}, M., {Reinsch}, K., {Thomas}, H., \& {Rampazzo}, R. 1992,
  \iaucirc, 5427, 1

\bibitem[{{Downes} \& {Duerbeck}(2000)}]{2000AJ....120.2007D}
{Downes}, R.~A., \& {Duerbeck}, H.~W. 2000, \aj, 120, 2007

\bibitem[{{Eyres} {et~al.}(2000){Eyres}, {Bode}, {O'Brien}, {Watson}, \&
  {Davis}}]{2000MNRAS.318.1086E}
{Eyres}, S.~P.~S., {Bode}, M.~F., {O'Brien}, T.~J., {Watson}, S.~K., \&
  {Davis}, R.~J. 2000, \mnras, 318, 1086

\bibitem[{{Finzell} {et~al.}(2017){Finzell}, {Chomiuk}, {Metzger}, {Walter},
  {Linford}, {Mukai}, {Nelson}, {Weston}, {Zheng}, {Sokoloski}, {Mioduszewski},
  {Rupen}, {Dong}, {Bohlsen}, {Buil}, {Prieto}, {Wagner}, {Bensby}, {Bond},
  {Sumi}, {Bennett}, {Abe}, {Koshimoto}, {Suzuki}, {P.}, {Tristram},
  {Christie}, {Natusch}, {McCormick}, {Yee}, \& {Gould}}]{2017arXiv170103094F}
{Finzell}, T., {Chomiuk}, L., {Metzger}, B.~D., {et~al.} 2017, ArXiv e-prints,
  arXiv:1701.03094

\bibitem[{{Gallagher} \& {Starrfield}(1976)}]{1976MNRAS.176...53G}
{Gallagher}, J.~S., \& {Starrfield}, S. 1976, \mnras, 176, 53

\bibitem[{{Gehrz} {et~al.}(1998){Gehrz}, {Truran}, {Williams}, \&
  {Starrfield}}]{1998PASP..110....3G}
{Gehrz}, R.~D., {Truran}, J.~W., {Williams}, R.~E., \& {Starrfield}, S. 1998,
  \pasp, 110, 3

\bibitem[{{Greisen}(2003)}]{2003ASSL..285..109G}
{Greisen}, E.~W. 2003, Information Handling in Astronomy - Historical Vistas,
  285, 109

\bibitem[{{Heywood} {et~al.}(2005){Heywood}, {O'Brien}, {Eyres}, {Bode}, \&
  {Davis}}]{2005MNRAS.362..469H}
{Heywood}, I., {O'Brien}, T.~J., {Eyres}, S.~P.~S., {Bode}, M.~F., \& {Davis},
  R.~J. 2005, \mnras, 362, 469

\bibitem[{{Hjellming}(1992)}]{1992IAUC.5473....3H}
{Hjellming}, R.~M. 1992, \iaucirc, 5473, 3

\bibitem[{{Hjellming}(1996)}]{1996ASPC...93..174H}
{Hjellming}, R.~M. 1996, in Astronomical Society of the Pacific Conference
  Series, Vol.~93, Radio Emission from the Stars and the Sun, ed. A.~R.
  {Taylor} \& J.~M. {Paredes}, 174

\bibitem[{{Hjellming} {et~al.}(1979){Hjellming}, {Wade}, {Vandenberg}, \&
  {Newell}}]{1979AJ.....84.1619H}
{Hjellming}, R.~M., {Wade}, C.~M., {Vandenberg}, N.~R., \& {Newell}, R.~T.
  1979, \aj, 84, 1619

\bibitem[{{Kasliwal} {et~al.}(2011){Kasliwal}, {Cenko}, {Kulkarni}, {Ofek},
  {Quimby}, \& {Rau}}]{2011ApJ...735...94K}
{Kasliwal}, M.~M., {Cenko}, S.~B., {Kulkarni}, S.~R., {et~al.} 2011, \apj, 735,
  94

\bibitem[{{Krauss} {et~al.}(2011){Krauss}, {Chomiuk}, {Rupen}, {Roy},
  {Mioduszewski}, {Sokoloski}, {Nelson}, {Mukai}, {Bode}, {Eyres}, \&
  {O'Brien}}]{2011ApJ...739L...6K}
{Krauss}, M.~I., {Chomiuk}, L., {Rupen}, M., {et~al.} 2011, \apjl, 739, L6

\bibitem[{{Kwok}(1983)}]{1983MNRAS.202.1149K}
{Kwok}, S. 1983, \mnras, 202, 1149

\bibitem[{{Linford} {et~al.}(2015){Linford}, {Ribeiro}, {Chomiuk}, {Nelson},
  {Sokoloski}, {Rupen}, {Mukai}, {O'Brien}, {Mioduszewski}, \&
  {Weston}}]{2015ApJ...805..136L}
{Linford}, J.~D., {Ribeiro}, V.~A.~R.~M., {Chomiuk}, L., {et~al.} 2015, \apj,
  805, 136

\bibitem[{{Mukai} \& {Ishida}(2001)}]{2001ApJ...551.1024M}
{Mukai}, K., \& {Ishida}, M. 2001, \apj, 551, 1024

\bibitem[{{Nelson} {et~al.}(2014){Nelson}, {Chomiuk}, {Roy}, {Sokoloski},
  {Mukai}, {Krauss}, {Mioduszewski}, {Rupen}, \&
  {Weston}}]{2014ApJ...785...78N}
{Nelson}, T., {Chomiuk}, L., {Roy}, N., {et~al.} 2014, \apj, 785, 78

\bibitem[{{Orio} {et~al.}(1996){Orio}, {Balman}, {della Valle}, {Gallagher}, \&
  {Oegelman}}]{1996ApJ...466..410O}
{Orio}, M., {Balman}, S., {della Valle}, M., {Gallagher}, J., \& {Oegelman}, H.
  1996, \apj, 466, 410

\bibitem[{{Pachoulakis} \& {Saizar}(1995)}]{1995ASSL..205..303P}
{Pachoulakis}, I., \& {Saizar}, P. 1995, in Astrophysics and Space Science
  Library, Vol. 205, Cataclysmic Variables, ed. A.~{Bianchini}, M.~{della
  Valle}, \& M.~{Orio}, 303

\bibitem[{{Prialnik}(1986)}]{1986ApJ...310..222P}
{Prialnik}, D. 1986, \apj, 310, 222

\bibitem[{{Ribeiro} {et~al.}(2014){Ribeiro}, {Chomiuk}, {Munari}, {Steffen},
  {Koning}, {O'Brien}, {Simon}, {Woudt}, \& {Bode}}]{Ribeiro}
{Ribeiro}, V.~A.~R.~M., {Chomiuk}, L., {Munari}, U., {et~al.} 2014, ArXiv
  e-prints, arXiv:1407.2935

\bibitem[{{Roy} {et~al.}(2012){Roy}, {Chomiuk}, {Sokoloski}, {Weston}, {Rupen},
  {Johnson}, {Krauss}, {Nelson}, {Mukai}, {Mioduszewski}, {Bode}, {Eyres}, \&
  {O'Brien}}]{2012BASI...40..293R}
{Roy}, N., {Chomiuk}, L., {Sokoloski}, J.~L., {et~al.} 2012, Bulletin of the
  Astronomical Society of India, 40, 293

\bibitem[{{Saizar} {et~al.}(1996){Saizar}, {Pachoulakis}, {Shore},
  {Starrfield}, {Williams}, {Rothschild}, \& {Sonneborn}}]{1996MNRAS.279..280S}
{Saizar}, P., {Pachoulakis}, I., {Shore}, S.~N., {et~al.} 1996, \mnras, 279,
  280

\bibitem[{{Sala} \& {Hernanz}(2005)}]{2005AA...439.1061S}
{Sala}, G., \& {Hernanz}, M. 2005, \aap, 439, 1061

\bibitem[{{Schwarz} {et~al.}(2011){Schwarz}, {Ness}, {Osborne}, {Page},
  {Evans}, {Beardmore}, {Walter}, {Helton}, {Woodward}, {Bode}, {Starrfield},
  \& {Drake}}]{2011ApJS..197...31S}
{Schwarz}, G.~J., {Ness}, J.-U., {Osborne}, J.~P., {et~al.} 2011, \apjs, 197,
  31

\bibitem[{{Seaquist} \& {Bode}(2008)}]{Seaquist_Bode08}
{Seaquist}, E.~R., \& {Bode}, M.~F. 2008, in {Classical Novae, 2nd
  Edition.~Cambridge Astrophysics Series, No.~43, Cambridge: Cambridge
  University Press}, ed. {M.~F.~Bode \& A.~Evans}, 141

\bibitem[{{Seaquist} {et~al.}(1980){Seaquist}, {Duric}, {Israel}, {Spoelstra},
  {Ulich}, \& {Gregory}}]{1980AJ.....85..283S}
{Seaquist}, E.~R., {Duric}, N., {Israel}, F.~P., {et~al.} 1980, \aj, 85, 283

\bibitem[{{Seaquist} \& {Palimaka}(1977)}]{1977ApJ...217..781S}
{Seaquist}, E.~R., \& {Palimaka}, J. 1977, \apj, 217, 781

\bibitem[{{Shepherd} {et~al.}(1994){Shepherd}, {Pearson}, \&
  {Taylor}}]{1994BAAS...26..987S}
{Shepherd}, M.~C., {Pearson}, T.~J., \& {Taylor}, G.~B. 1994, in Bulletin of
  the American Astronomical Society, Vol.~26, Bulletin of the American
  Astronomical Society, 987--989

\bibitem[{{Sonneborn} {et~al.}(1992){Sonneborn}, {Shore}, \&
  {Starrfield}}]{1992IAUC.5428....2S}
{Sonneborn}, G., {Shore}, S.~N., \& {Starrfield}, S.~G. 1992, \iaucirc, 5428, 2

\bibitem[{{Starrfield} {et~al.}(2008){Starrfield}, {Iliadis}, \&
  W}]{Starrfield}
{Starrfield}, S., {Iliadis}, C., \& W, H.~R. 2008, in {Classical Novae, 2nd
  Edition.~Cambridge Astrophysics Series, No.~43, Cambridge: Cambridge
  University Press}, ed. {M.~F.~Bode \& A.~Evans}, 77

\bibitem[{{Starrfield} {et~al.}(2000){Starrfield}, {Truran}, \&
  {Sparks}}]{2000NewAR..44...81S}
{Starrfield}, S., {Truran}, J.~W., \& {Sparks}, W.~M. 2000, New Ast. Rev., 44,
  81

\bibitem[{{Starrfield} {et~al.}(1998){Starrfield}, {Truran}, {Wiescher}, \&
  {Sparks}}]{1998MNRAS.296..502S}
{Starrfield}, S., {Truran}, J.~W., {Wiescher}, M.~C., \& {Sparks}, W.~M. 1998,
  \mnras, 296, 502

\bibitem[{{Stickland} {et~al.}(1981){Stickland}, {Penn}, {Seaton}, {Snijders},
  \& {Storey}}]{1981MNRAS.197..107S}
{Stickland}, D.~J., {Penn}, C.~J., {Seaton}, M.~J., {Snijders}, M.~A.~J., \&
  {Storey}, P.~J. 1981, \mnras, 197, 107

\bibitem[{{Taylor} {et~al.}(1987){Taylor}, {Pottasch}, {Seaquist}, \&
  {Hollis}}]{1987A&A...183...38T}
{Taylor}, A.~R., {Pottasch}, S.~R., {Seaquist}, E.~R., \& {Hollis}, J.~M. 1987,
  \aap, 183, 38

\bibitem[{{Townsley} \& {Bildsten}(2004)}]{2004RMxAC..20..192T}
{Townsley}, D.~M., \& {Bildsten}, L. 2004, in Revista Mexicana de Astronomia y
  Astrofisica Conference Series, Vol.~20, Revista Mexicana de Astronomia y
  Astrofisica Conference Series, ed. G.~{Tovmassian} \& E.~{Sion}, 192--193

\bibitem[{{Wade} {et~al.}(2000){Wade}, {Harlow}, \&
  {Ciardullo}}]{2000PASP..112..614W}
{Wade}, R.~A., {Harlow}, J.~J.~B., \& {Ciardullo}, R. 2000, \pasp, 112, 614

\bibitem[{{Weston} {et~al.}(2015{\natexlab{a}}){Weston}, {Sokoloski},
  {Metzger}, {Zheng}, {Chomiuk}, {Krauss}, {Linford}, {Nelson}, {Mioduszewski},
  {Rupen}, {Finzell}, \& {Mukai}}]{2015arXiv150505879W}
{Weston}, J.~H.~S., {Sokoloski}, J.~L., {Metzger}, B.~D., {et~al.}
  2015{\natexlab{a}}, ArXiv 1505.05879, arXiv:1505.05879

\bibitem[{{Weston} {et~al.}(2015{\natexlab{b}}){Weston}, {Sokoloski},
  {Chomiuk}, {Linford}, {Nelson}, {Mukai}, {Finzell}, {Mioduszewski}, {Rupen},
  \& {Walter}}]{2015arXiv151006751W}
{Weston}, J.~H.~S., {Sokoloski}, J.~L., {Chomiuk}, L., {et~al.}
  2015{\natexlab{b}}, arXiv 1510.06751, arXiv:1510.06751

\bibitem[{{Weston} {et~al.}(2016){Weston}, {Sokoloski}, {Metzger}, {Zheng},
  {Chomiuk}, {Krauss}, {Linford}, {Nelson}, {Mioduszewski}, {Rupen}, {Finzell},
  \& {Mukai}}]{2016MNRAS.457..887W}
{Weston}, J.~H.~S., {Sokoloski}, J.~L., {Metzger}, B.~D., {et~al.} 2016,
  \mnras, 457, 887

\bibitem[{{Wilkes} {et~al.}(1981){Wilkes}, {Ferland}, {Hanes}, \&
  {Truran}}]{1981MNRAS.197....1W}
{Wilkes}, B.~J., {Ferland}, G.~J., {Hanes}, D., \& {Truran}, J.~W. 1981,
  \mnras, 197, 1

\bibitem[{{Williams} {et~al.}(1994){Williams}, {Phillips}, \&
  {Hamuy}}]{1994ApJS...90..297W}
{Williams}, R.~E., {Phillips}, M.~M., \& {Hamuy}, M. 1994, \apjs, 90, 297

\bibitem[{{Wolf} {et~al.}(2013){Wolf}, {Bildsten}, {Brooks}, \&
  {Paxton}}]{2013ApJ...777..136W}
{Wolf}, W.~M., {Bildsten}, L., {Brooks}, J., \& {Paxton}, B. 2013, \apj, 777,
  136

\bibitem[{{Woudt} \& {Warner}(2001)}]{2001MNRAS.328..159W}
{Woudt}, P.~A., \& {Warner}, B. 2001, \mnras, 328, 159

\bibitem[{{Woudt} {et~al.}(2009){Woudt}, {Steeghs}, {Karovska}, {Warner},
  {Groot}, {Nelemans}, {Roelofs}, {Marsh}, {Nagayama}, {Smits}, \&
  {O'Brien}}]{2009ApJ...706..738W}
{Woudt}, P.~A., {Steeghs}, D., {Karovska}, M., {et~al.} 2009, \apj, 706, 738

\bibitem[{{Yaron} {et~al.}(2005){Yaron}, {Prialnik}, {Shara}, \&
  {Kovetz}}]{2005ApJ...623..398Y}
{Yaron}, O., {Prialnik}, D., {Shara}, M.~M., \& {Kovetz}, A. 2005, \apj, 623,
  398

\end{thebibliography}

\end{document}